\documentclass[twocolumn,eqsecnum,preprintnumbers,nofootinbib]{revtex4}

\usepackage{amsmath,amssymb,amsfonts,times,graphicx}
\usepackage[ps2pdf,bookmarks=true,colorlinks,linkcolor=red,urlcolor=blue,citecolor=blue]{hyperref}
\usepackage[usenames]{color}
\usepackage{dcolumn}

\usepackage[normalem]{ulem}
\usepackage{enumerate}
\usepackage{epsfig}
\usepackage{yfonts}

\usepackage{bm}

\def\eg{{\it e.g. }}
\def\ie{{\it i.e.}}

\def\({\left(}
\def\){\right)}
\def\[{\left[}
\def\]{\right]}
\def\<{\langle}
\def\>{\rangle}

\def\CT{{\cal T}}
\def\CM{{\cal M}}

\def\CO{{\cal O}}

\def\Dslash{\rlap{\hskip0.3em/}D}

\newcommand\half{{\ensuremath{\frac{1}{2}}}}

\newcommand\vev[1]{{\ensuremath{\left\langle{#1}\right\rangle}}}

\def\Ione{\hbox{$1\hskip -1.2pt\vrule depth 0pt height 1.53ex width 0.7pt
                  \vrule depth 0pt height 0.3pt width 0.12em$}}

\newcommand{\be}{\begin{equation}}
\newcommand{\ee}{\end{equation}}
\newcommand{\bea}{\begin{eqnarray}}
\newcommand{\eea}{\end{eqnarray}}
\newcommand{\bwt}{\begin{widetext}}
\newcommand{\ewt}{\end{widetext}}

\newcommand{\bi}{\begin{itemize}}
\newcommand{\ei}{\end{itemize}}
\newcommand{\ben}{\begin{enumerate}}
\newcommand{\een}{\end{enumerate}}
\newcommand{\bca}{\begin{cases}}
\newcommand{\eca}{\end{cases}}
\newcommand{\bln}{\begin{align}}
\newcommand{\eln}{\end{align}}
\newcommand{\bst}{\begin{split}}
\newcommand{\est}{\end{split}}

\begin{document}

\title{Non-Abelian statistics versus the Witten anomaly}

\preprint{MIT-CTP/4266}

\author{John McGreevy and Brian Swingle}

\affiliation{Department of Physics,
Massachusetts
Institute of Technology,
Cambridge, MA 02139 }

\begin{abstract}

This paper is motivated by prospects for non-Abelian statistics of deconfined particle-like objects in 3+1 dimensions,
realized as solitons with localized Majorana zeromodes.
To this end, we study the fermionic collective coordinates of magnetic monopoles in 3+1 dimensional
spontaneously-broken SU$(2)$ gauge theories with various spectra of fermions.
We argue that
a single Majorana zeromode of the monopole
is not compatible with cancellation of the Witten SU$(2)$ anomaly.
We also compare this approach with other attempts to realize deconfined non-Abelian objects in $3+1$ dimensions.

\end{abstract}

August, 2011

\maketitle
\vspace{10mm}\smash{\raisebox{29mm}{\makebox[10cm][c]{\hspace{26cm}\includegraphics[height=2.6cm]{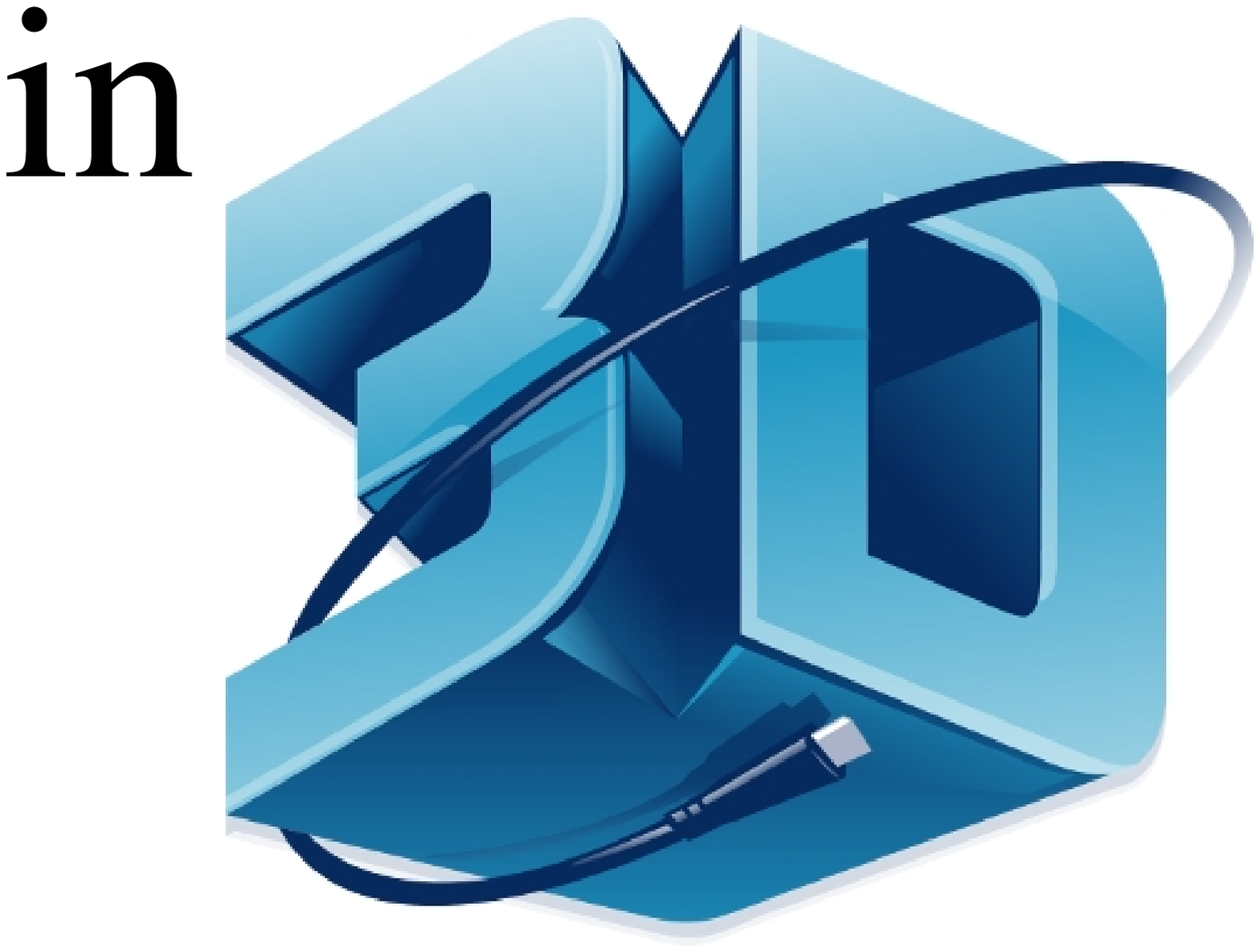}}}}

\tableofcontents

\section{Introduction}

Point particles
in 3+1 dimensions cannot have non-Abelian statistics
because of the triviality of the topology of their configuration space
\cite{Leinaas:1977fm}.
However, a particle-like object with extra structure can
have a configuration space with more interesting homotopy.
Inspired by ideas from topological insulators \cite{Fu:2008zzb},
Teo and Kane \cite{Teo:2009qv} recently made a specific proposal in this direction.
The objects in question are hedgehogs of a 3-component
order parameter, coupled
to fermionic excitations that are gapped in the presence of a non-zero order parameter.
Freedman {\it et al} \cite{Freedman:2010ak} show that these objects exhibit what they call {\it projective ribbon statistics};
the data needed to specify a configuration include the preimage under the order parameter map of the north pole
and a nearby point.

The hedgehog defects support
real fermionic zero modes and multiple hedgehogs are associated with a
non-local Hilbert space generated by the zero mode operators. Motions
of the hedgehogs implement unitary transformations in the non-local
Hilbert space, a concept familiar from topological quantum computing
in $2+1$ dimensions \cite{Nayak:2008zza}.  Because exchanging identical particles leads to a
non-trivial unitary transformation of the quantum state instead of
merely a phase, we say such objects have non-Abelian statistics.  The
presence of Majorana zero modes and the non-trivial configuration
space are both crucial to this story.

Freedman {\it et al} also point out the following problematic feature of the model of Teo and Kane:
if the order parameter field has a nonzero stiffness,
a single hedgehog is not a finite-energy configuration.
Configurations with zero net hedgehog number can have finite energy,
but there will
be a confining force between the hedgehogs due to
gradient energy in the order parameter field.  This energy cost will scale at least linearly with the separation between the hedgehogs. The cost may be even higher in the absence of full SU$(2)$ symmetry for the order parameter (and such symmetry is unlikely given that the order parameter involves both superconducting and particle-number-conserving terms).
This makes it difficult to imagine adiabatically moving these solitons
around each other.

Putting aside the issues
with using this proposal as a platform for quantum computing
(note further that braiding of Majoranas does not provide a set of universal gates),
we would like to confront the conceptual question of whether it is possible in principle to deconfine such non-Abelian particles in $3+1$ dimensions.  We are also interested more generally in what happens to Majorana zero modes when the relevant order parameter field begins to fluctuate.
If we were able to deconfine non-Abelian particles in $3+1$ dimensions, there would be profound practical and conceptual implications.

One suggestion for removing the confining energy
follows the analogous step in the study of
vortices in 2d: gauge the rotation symmetry in the
order parameter space.
If all directions are gauge-equivalent, there
need not be a confining energy between the hedgehogs,
which in the resulting gauge theory are 't Hooft-Polyakov monopoles \cite{'tHooft:1974qc,Polyakov:1974ek}.
(There will be a magnetic Coulomb force between the monopoles,
but this falls off with their separation.)
But 3+1 dimensional SU$(2)$ gauge theory with
the requisite fermion content,
namely a single Weyl doublet (\ie\ eight
Majorana fields),
suffers from the Witten SU$(2)$ anomaly \cite{Witten:1982fp}
(as \cite{Freedman:2010ak} also observe).
One implication of this is that the gauge field partition sum
vanishes identically.
Another pathology resulting from the anomaly
is a violation of fermion parity by the gauge dynamics.  Specifically, an instanton creates a single fermion in violation of fermion parity.
The addition of an adjoint Higgs field
(relative to the discussion of \cite{Witten:1982fp})
doesn't change the structure of the fermion
determinant which is responsible for the fatal factor of $-1$ (which it acquires
under the gauge transformations which represent the nontrivial element of $\pi_4({\rm SU}(2))$),
as we argue below in \S\ref{sec:persistence}.

We will construct below a microscopically-consistent theory which, in a range of energy scales,
looks like this Witten-anomalous SU$(2)$ gauge theory in the Higgs phase
with a single Weyl doublet.
The spectrum of fermionic particles with $E < M_W$ is identical to that of the theory described above;
at these energies, the Witten anomaly is cancelled by a certain
Wess-Zumino-Witten term made from the adjoint scalar and the gauge field.
(This situation is similar, but not identical, to models discussed by
d'Hoker and Farhi \cite{D'Hoker:1983kr, D'Hoker:1984ph}.)
However, this term is ill-defined when the order parameter vanishes,
as it does in the core of the monopole, and
we must provide a UV completion to address the question
of whether the monopole has a Majorana zeromode. The simplest UV completion of this model
involves adding in another Weyl fermion doublet.

Before preceeding with an analysis of the SU$(2)$ gauge theory,
we pause to consider an alternative possible route to deconfine the localized objects hosting Majorana modes.  Instead of gauging the SU$(2)$ symmetry that is spontaneously broken by the order parameter field in the Teo-Kane model, we can consider disordering the broken phase into a liquid-like phase without any broken symmetry
\cite{Freedman:2010ak, Senthil:discussions}.  Importantly, we must  achieve this disordering without proliferating the monopole defects that hosted Majorana modes, otherwise we will trivially lose the localized Majorana mode.
The simplest-to-describe disordered phase has a description in terms of an emergent U$(1)$ gauge field, and the hedgehog defects, assuming they have finite energy, become magnetic monopoles in the U$(1)$ gauge theory:
we are again led to a description in terms of magnetic charges in an Abelian gauge theory.
(We describe other possibilities for disordered phases in the last section.)
Now the important question is:
do the Majorana modes survive the disordering process, and if so, are the magnetic monopoles in this theory deconfined (only interacting via a long range Coulomb interaction) particles carrying Majorana zero modes?
Again the question of the survival of the Majorana modes requires short-distance information
about the theory.
Later we will return to this question for the disordered state,
arguing on general grounds that this particular scenario is unlikely.

To clarify, our desiderata for deconfined non-Abelian excitations in 3+1 dimensions are as follows.
First we will discuss the desired form of the regulated theory at high energies and then the form of the theory at low energies.  From the point of view of condensed matter physics, we would most like to have a microscopic lattice model involving only spin-like or electron-like degrees of freedom that enters a phase where there are deconfined particles.  We do not accept as a valid realization a model that contains Majorana fermion degrees of freedom in a microscopic lattice model.  We make this requirement because we do not want to put the Majoranas in ``by hand".  However, we would permit a Majorana based lattice model provided we could reinterpret it as an intermediate scale description arising from a truly microscopic model of electrons, likely in the presence of superconductivity (a bipartite lattice is a sufficient condition).  From the point of view of high energy physics, we would like to have an anomaly free gauge theory coupled to fermions and scalars that has a non-perturbative regularization of some type, be it lattice gauge theory or string theory.  In the high energy way of thinking, we do not require the absence of gauge fields in the microscopic description, for example, we would accept an asymptotically free gauge theory interacting with Dirac fermions.

In the low energy theory, we have two general interests.  First, any putative non-Abelian particle-like excitations should have a clearly defined configuration space.  We should have a clear understanding of the non-locality inherent in this configuration space that permits otherwise point-like objects to have interesting statistics.   Second, it must be possible to perform motions of the non-Abelian excitations without high energy cost, without dramatically exciting other degrees of freedom, without violating causality or unitarity, and without producing decoherence in the space of ``protected" states.  For example, decoherence due to unscreened gauge fields limits our ability to superpose states with macroscopically different charge configurations.  We emphasize especially the issue of the low energy configuration space.  This space must be rich enough to support representations of its fundamental group that are non-local, as with non-Abelian anyon representations of the braid group in $2+1$.  The symmetric group is known to be insufficient for this purpose \cite{Doplicher:1971wk,Doplicher:1973at}, and indeed as a finite group its image in any unitary group must be quite limited.

In this paper we study the possibility of non-Abelian particle-like excitations in a
3+1 dimensional field theory. In particular, we explore the apparent conflict between
a single Majorana zeromode of the 't Hooft-Polyakov monopole (we will refer to such an object as a `Majorana monopole')
and microscopic consistency of the SU$(2)$ gauge theory.

The outline of the paper is as follows.
In the next section we
generalize the classic analysis of Jackiw and Rebbi
\cite{Jackiw:1975fn}
to construct the zeromode solution
of the Dirac equation in the Witten-anomalous theory described above.
In section III, we discuss the cancellation of the Witten anomaly
and its effects on the zeromode structure of the monopole.  In section IV
we discuss an instructive example in 4+1 dimensions.
In section V we provide general arguments for
obstructions to Majorana monopoles in 3+1 dimensions following the desiderata described above.

Related work appears in \cite{qpaper}, which studies an interesting fermion dimer model whose low energy physics includes majorana monopoles interacting with gauge fields as well as gapless fermionic degrees of freedom. Some features of 3d non-Abelian particles appear to be realized in their model, but we emphasize that their conclusions do not contradict our own; our analysis suggests that the gapless fermions are essential.  \cite{qpaper} also studies a 5d model similar to the one discussed in \S\ref{sec:5d}.

\section{Majorana monopoles in an anomalous theory}

Consider an SU$(2)$ gauge theory in 3+1 dimensions
with a scalar field $\Phi$ in the adjoint representation;
we will suppose that the action for $\Phi$ is such
that in the ground state it breaks SU$(2)$ down to U$(1)$.
Include also a single SU$(2)$ doublet of Weyl fermions, $\chi$;
altogether there are $2^3 = 8$ real fermion degrees of freedom.
This is half as many fermion degrees of freedom as considered by Jackiw and Rebbi
in their 3+1-d discussion
\cite{Jackiw:1975fn}, and the same number as considered by Witten \cite{Witten:1982fp}.
As we demonstrate next, this theory suffers from the Witten anomaly
-- if we try to quantize the gauge field,
we get nonsense.  Specifically, the partition function vanishes and expectation values of gauge invariant observables are undefined.
For the discussion in this section the bosonic fields $\Phi, A$ will therefore be treated as background fields.

Consider the fermion Lagrange density
\be
\label{eq:fermL}
L_{\rm fermions} =
\chi^{\dagger} i \bar \sigma^\mu D_\mu \chi -
\half \lambda \chi^T i \sigma^2 i \tau^2 \vec \tau \cdot \vec \Phi \chi + h.c.
\ee
Here $\chi_{\alpha a} $ is a (left-handed) Weyl doublet of SU$(2)$:
$\alpha = 1,2$ is a spin index,   $ a = 1,2$ is a gauge index.
$\bar \sigma^\mu  = (1, - \vec \sigma)^\mu$.
The covariant derivative is defined as
$ \( D_\mu \chi\)_{\alpha a} =  \partial_\mu \chi_{\alpha a} - i g A_{\mu ab} \chi_{\alpha b} $
where $A_{\mu ab} $ is the SU$(2)$ gauge field.
$\lambda$ is a complex coupling constant.
Note that because $\chi$ is in a pseudoreal representation of
both the SU$(2)$ gauge group
and the Lorentz group, the
object
$\chi^C \equiv \chi^T i \sigma^2 i \tau^2 $
transforms in the conjugate representation of both groups.
There is no nonvanishing, gauge-invariant and Lorentz-invariant mass term
(not involving the Higgs field $\Phi$)
with this field content.
We will comment in
\S\ref{sec:NRmass}
on the effect of Lorentz-breaking terms of the form $\chi^\dagger \chi$.

This theory has two independent mass scales:
the mass of the $W$-bosons, $m_W = g v$
($v$ is the vev of the adjoint Higgs field, $g$ is the SU$(2)$ gauge coupling at the
scale $m_W$),
and the mass of the fermion $ \lambda v$.

\subsection{Persistence of Witten anomaly}
\label{sec:persistence}

The addition of the adjoint scalar $\Phi$
and its coupling to the fermion doublet
does not modify the
anomalous transformation law of the fermion determinant.
That this is the case can be seen by embedding the theory in an SU$(3)$ gauge theory
with a perturbative gauge anomaly as in \cite{Witten:1983tx,Elitzur:1984kr,Klinkhamer:1990eb}.
The relevant theory has an SU$(3)$ adjoint scalar $\tilde \Phi$,
an SU$(3)$ triplet of Weyl fermions $\tilde \chi$ and an SU$(3)$ triplet of scalars $\Upsilon$,
with the coupling
\be
\label{eq:su3}
L_{\text{SU}(3)} \supset \tilde \chi_a^T i \sigma^2 \Upsilon_b \epsilon_{abc} \tilde \Phi_{cd} \tilde \chi_d ,
\ee
where $a=1,2,3$ is a triplet index.
Condensing the scalar triplet $\vev{\Upsilon} = \lambda$ breaks
the SU$(3)$ down to SU$(2)$, and
the coupling
\eqref{eq:su3}
reduces to the desired coupling
between the Weyl fermions charged under the unbroken SU$(2)$ and
the adjoint scalar in \eqref{eq:fermL}.  The form of the perturbative SU$(3)$ anomaly
is unaffected by the addition of scalars
and so the calculation of the variation of the fermion measure
by integrating the SU$(3)$ anomaly
\cite{Witten:1983tx,Elitzur:1984kr,Klinkhamer:1990eb}
is unmodified compared to the theory without scalar fields.

\subsection{The Majorana zeromode}

The Dirac equation which results from varying
\eqref{eq:fermL} is
\be\label{eq:dirac1}
 0 = \delta_{\bar \chi} S_{{\rm fermion}}
 =-  i \bar \sigma^\mu D_\mu \chi
+ \lambda^\dagger  i \sigma^2 \Phi \cdot \tau i \tau^2  \chi^\star~.
 \ee
We consider this Dirac equation
in the background of the
't Hooft-Polyakov monopole solution,
\be
A^B_0 = 0~; ~~~~A_i^B = \epsilon_{ijB} \hat r^j A(r)~;
~~~~ \Phi^B = \hat r ^B \phi(r)
\ee
($B = 1,2,3$ is an adjoint index)
with
\be\label{eq:falloff} A(r) \buildrel{r \to \infty}\over{\approx} 1/r,
~~~\phi(r)  \buildrel{r \to \infty}\over{\approx} v .\ee

A zero-energy solution of \eqref{eq:dirac1} is of the form:
$ \chi_{\alpha a} = i \tau^2_{\alpha a} g(r)$
(where $\alpha$ is the spin index and $a$ is the SU$(2)$ index).
This is the same ansatz as in equation A4 of \cite{Jackiw:1975fn}.
With this substitution, the zeromode equation
reduces to
\be(\partial_i + 2 \hat r_i A ) g + i \lambda \phi \hat r_i g^\star = 0 .\ee
By rephasing the $\chi$ field, we can assume WLOG that $\lambda$ is real and positive.
The solution for $g$ is then
\be \label{eq:majoranazm}
g(r)= c e^{- \pi i /4 } e^{ - \int^r ( \lambda \phi - 2 A ) } \ee
where $c$ is a {\it real} constant.
We emphasize that the phase of the normalizable solution is determined by normalizability of the solution at large $r$.

Quantizing this fermionic collective coordinate gives
a Majorana fermion acting on the monopole Hilbert space,
which is represented by a unique state.
This leads inevitably to non-Abelian statistics for the monopoles,
in the same manner as expected for vortices in p+ip superconductors
or the pfaffian quantum Hall state
\cite{Nayak:1996,Read:2000,Ivanov:2001,Sato:2003iz,Nayak:2008zza}.
Briefly, two widely-separated monopoles will have two Majorana
zeromodes, which can be combined into $ c = \frac{\gamma_1 + i \gamma_2}{2}$,
with $\{c, c^\dagger\} = 1$;
this algebra must be represented by a two-state system.
Interchanging the monopoles adiabatically implements the operator $$U_{1 \rightleftharpoons 2} = \exp{\(\frac{\pi}{4} \gamma_1 \gamma_2 \)} = \exp{\( i \frac{\pi}{4}(1 - 2 c^+ c)\)} $$
With two pairs of monopoles we could perform operations which
do not commute with each other.

We note that the coupling of $\chi$ to the gauge field
does not play a crucial role in generating this zeromode;
since
(by \eqref{eq:falloff})
the dominant term in the exponent of
\eqref{eq:majoranazm} at large $r$ comes from the scalar profile,
the gauge field can be set to zero without interfering with the zeromode.
The existence of the zermode solution without the gauge field
essentially follows from the analysis of \cite{Teo:2009qv}.

If only this were a consistent quantum system.  We describe one pathology of this system.  Recall that in the case of a Dirac fermion there is a complex fermion zero mode in the ungauged theory.  Once the SU$(2)$ symmetry is gauged, the low energy gauge group is U$(1)$ and hedgehog configurations become magnetic monopoles.  Now what happens to the two states living on a hedgehog in the ungauged theory?  In fact \cite{Jackiw:1975fn}, they become bosonic, having charge $\pm 1/2$ under the unbroken U$(1)$ due to the low energy U$(1)$ theta term of $\pi$.  To see this, assume that the charge $-1/2$ state is bosonic, then when we add a fermion in the zero mode of charge $1$ we reach a state of charge $1/2$ which would appear to differ in spin by $1/2$ from the bosonic state.  But we have forgotten the gauge field which adds extra angular momentum.  Indeed, a unit charge orbiting a minimal monopole leads to a gauge field configuration with angular momentum given by a half integer.  This extra half integer angular momentum when combined with the bare half integer angular momentum of the fermion leads again to a bosonic state.  In fact, we can check from the structure of the zero mode that the position and spin of the fermion are correlated so that no matter where the fermion is measured its spin will always compensate the $\hat{r}$ angular momentum contribution coming from the field.

Now the puzzle: in the case of a single Weyl fermion, we found that the complex fermion was replaced by a real zero mode, but what should happen when we turn on the gauge field?  Heuristically, we should obtain half of the pair of states with charges $\pm 1/2$.  Let $\CO_{+1}$ be an operator that moves us from the $-1/2$ to the $1/2$ charge state so that $\CO_{+1}$ carries charge 1.  By analogy with the definition of the Majorana fermion, an apparently interesting combination to consider is $\CO_{+1} + \CO_{+1}^{\dagger}$, but this operator creates states that decohere in the presence of the fluctuating U$(1)$ field;
we can identify no candidate for the pointer states into which they should decohere.
Is the Witten anomaly to blame?  The simplest resolution of the Witten anomaly, namely adding a second identical Weyl doublet, removes the spectre of decoherence by adding an extra real zero mode in the monopole core allowing for complex solutions, as we'll see next.

To summarize, we found an SU$(2)$ gauge theory where magnetic monopoles of an unbroken U$(1)$ gauge field appear to carry Majorana zero modes.  However, this theory suffers from the Witten anomaly rendering all gauge invariant observables ill-defined.  Related pathologies include a violation of fermion number by instantons and decoherent U$(1)$ charge superpositions.  In what follows, we try to cure the Witten anomaly while preserving the zero mode structure of the monopole.

\section{Cancelling the Witten anomaly}

It is possible to cancel the Witten anomaly by adding to the action
a certain functional of the adjoint scalar.
To see that this is the case, consider
integrating out a Weyl fermion $\chi_2$ coupled to the scalar field as above:
\be
\label{eq:integratechi}
\ln \int D\chi_2
\exp\(i S_{\text{ferm}}[\chi_2] \)
= \Gamma[\Phi, A] + {\text{non-universal stuff}}.
\ee
The functional $\Gamma$ defined by this equation 
is well-behaved because of 
the gap in the fermion spectrum.
The both-hand side of equation \eqref{eq:integratechi} must shift by $\pi$ (mod $2\pi$)
under an SU$(2)$ gauge transformation representing the nontrivial
class of
$\pi_4({\rm SU}(2))$.
The fact that the non-universal, short-distance stuff on the RHS does not accomplish
this shift follows because it is not sensitive to the topology of spacetime.

It is difficult to give an explicit expression for the functional $\Gamma$.
Naively, the WZW term for SU$(2)$ vanishes identically.
However, our term is not quite the usual WZW term since it arises from a
pfaffian rather than a determinant,
\ie\ it is invariant under only real-linear basis changes.
A similar situation
with different fermion representations
arises in \cite{D'Hoker:1983kr, D'Hoker:1984ph},
where
the effective action contains terms taking the form of
the gauge variation of our functional $\Gamma$.
In \S\ref{sec:persistence}, we have determined
the anomalous transformation of $\Gamma$ by embedding
into a theory with a perturbative anomaly;
this trick does not immediately determine the form of $\Gamma$ itself.
It would be useful to find an explicit expression for this functional.

One thing about $\Gamma$, however, is certain: it is ill-defined when
the order parameter $\Phi$ is not invertible.
A simple argument for this is that only when $\Phi$
is invertible are the fermion degrees of freedom gapped.
Therefore, in any field configuration where $\Phi$ vanishes,
such as the core of the magnetic monopole,
a model where the Witten anomaly is cancelled
by the variation of $\Gamma[\Phi]$ requires a
UV completion.

The simplest way to do this is obviously
to integrate in the second Weyl doublet $\chi_2$
by which we proved the existence of $\Gamma$;
we study this possibility next.
Are there other ways?
In the final section, we
will argue that the answer is `no'.

\subsection{Generic couplings in the two-Weyl-doublet theory}

Consider the fermion lagrangian density
\bea\label{eq:twoweyl}
L_{\rm 2 fermions} &=&
\chi^{I\dagger} i \bar \sigma^\mu D_\mu \chi_I -
{1\over 2} \lambda^{IJ} \chi_I^T i \sigma^2 i \tau^2 \vec \tau \cdot \vec \Phi \chi_J + h.c.
\cr\cr &&
- {1\over 2} m^{IJ} \chi_I^T i \sigma^2 i \tau^2 \chi_J + h.c.~~~
\eea
Here $\chi_{I\alpha a} $ are a pair of (left-handed) Weyl doublets of SU$(2)$:
$ I= 1,2$ is a flavor index, $\alpha = 1,2$ is a spin index,   $ a = 1,2$ is a gauge index.
Altogether there are now $2^3 = 8$ complex fermion degrees of freedom.
This is the same set of fermion degrees of freedom considered by Jackiw and Rebbi
and twice as many as considered by Witten.

We now comment on symmetries of this action, and
simplifications that can be made by field redefinitions of the fermions.
The Yukawa coupling term is more explicitly written as 
\bea
\label{lambdaterm}
&&\lambda^{IJ} \chi_I^T i \sigma^2 i \tau^2 \vec \tau \cdot \vec \Phi \chi_J + h.c. = \cr
\nonumber
&& \lambda^{IJ} \chi_I^T i \sigma^2 i \tau^2 \vec \tau \cdot \vec \Phi \chi_J +
\lambda^{\dagger~IJ} \chi_I^\dagger \vec \tau \cdot \vec \Phi i \sigma^2 i \tau^2  \chi^\star_J~.
\eea
The matrix $\lambda$ is symmetric, $\lambda^{IJ} = \lambda^{JI}$ by Fermi statistics.
A general complex symmetric matrix is not diagonalizable,
but rather has different right and left eigenvalues.
A complex symmetric matrix has a
singular value decomposition
(SVD) (called Takagi decomposition) of the form
\be \lambda = W d W^T \ee
where $d$ is diagonal with real, positive entries
$d = \( \begin{matrix} \lambda_1 & 0 \cr 0 & \lambda_2
\end{matrix}\) $, and $W$ is unitary.

Rephasing the fermion fields by a unitary rotation $U = (U^{- 1})^{\dagger}$
\be\label{eq:rephasing} \chi_I \to U_I^J \chi_J
\ee
changes the coupling matrix $\lambda$ by
\be \lambda \to  U \lambda U^T = U W d W^T U^T .\ee
Choosing $ U = W^{-1}$ gives $\lambda = d$.

By Fermi statistics, the Dirac mass matrix $m_{IJ}$ is antisymmetric, $m_{IJ} = m \epsilon_{IJ}$.
The effect of the rephasing \eqref{eq:rephasing}
on the Dirac mass is therefore
$ m_{IJ} \to  m_{IJ} \det U $.
Having fixed our freedom to rephase the fermions, the phase of the Dirac mass $m$ will be significant.  Global symmetries can constrain the phase of $m$.
In particular, with a Hermitian mass matrix, $m = m^\dagger$,
the model preserves a CP symmetry
 which acts by
$ \chi \mapsto i \sigma^2 i \tau^2 \chi^\star$.

When $\lambda$ is purely off-diagonal $\lambda = \( \begin{matrix} 0 & \lambda_0 \cr \lambda_0 & 0
\end{matrix}\) $
the system admits an extra U$(1)$ symmetry under which
\be \chi_1 \mapsto e^{ i \theta} \chi_1, ~~~
\chi_2 \mapsto e^{-  i \theta} \chi_2 ~.\ee
When the Dirac mass vanishes, the resulting
model is identical to the model studied in \cite{Jackiw:1975fn}.
To see this, construct from the two left-handed Weyl doublets
a single Dirac fermion
\be
\Psi \equiv \( \begin{matrix} \chi_1 \cr \chi_2^\star i \tau^2 i \sigma^2 \end{matrix} \).
\ee
Then the action \eqref{eq:twoweyl}, with $\lambda$
off-diagonal and $\lambda_0\equiv\lambda_0^R + i \lambda_0^I$, is
\be
 L_{\rm 2 fermions} =
 \bar \Psi i \Dslash \Psi -  \bar \Psi \(\lambda_0^R + i \lambda_0^I \gamma^5\)\vec \tau \cdot \vec \Phi \Psi
 + m \bar \Psi \Psi.
\ee
Returning to the SVD form of the action, this is equivalent to the case where
the diagonal entries are equal
$\lambda_1 = \lambda_2 $.  In this basis, the U$(1)$ symmetry acts as the SO(2) rotation
\be \chi_1 + i \chi_2 \mapsto e^{  i \theta}  \(  \chi_1  +i \chi_2\) .\ee

The general two-Weyl-doublet theory now has three mass scales:
the mass of the $W$-bosons, $m_W = g v$,
and the masses of the two Weyl fermions $ \lambda_{1,2} v$.
In the regime
\be  \lambda_1 v \ll m_W \ll \lambda_2 v \ee
we have a large window of energies
in which the bulk spectrum
is that of the Witten-anomalous theory studied above.

We note that the theory with two Weyl doublets admits
a Lorentz-violating (but gauge-invariant and rotation-invariant) mass term
of the form
\be L_{NR} = M^{IJ} \chi^\dagger_I \chi_J .\ee
We will comment below in \S\ref{sec:NRmass} on its effects on the zeromode structure.

\subsection{FZMs in the two-Weyl-doublet theory}

The Dirac equation is now
\bea
 0 &=& \delta_{\bar \chi_I} S_{{\rm fermion}}
 =-  i \bar \sigma^\mu D_\mu \chi_I
 \cr &&
+ \lambda^\dagger_{IJ}  i \sigma^2 \Phi \cdot \tau i \tau^2  \chi^\star_J
+ m^\dagger_{IJ} i \sigma^2  i \tau^2 \chi_J
 \eea
When the Dirac mass $m=0$, in the basis where $\lambda$ is diagonal,
the zeromode equations for $\chi_{1,2}$ decouple,
and each is of the form of \eqref{eq:majoranazm}.
There are then two real solutions:
\be
\chi_{I\alpha a}(r)=  i \tau^2_{\alpha a} g_I ,
~~~ g_I = c_I e^{- \pi i /4 } e^{ - \int^r ( \lambda_I \phi - 2 A ) }~.
\ee

\subsection{Effect of the Dirac mass}

With a nonzero Dirac mass, the zeromode equations for $\chi_{1,2}$ are coupled.
A nonzero Dirac mass requires any putative zeromode solution to
include also a triplet component, {\it i.e.} to have the more general form
\be \chi_{a \alpha I} = i \tau^2_{a \alpha} g_I + i \( \tau^2 \tau^i \)_{a \alpha} g^i_I .\ee
The zero-energy Dirac equation is
\be
0 = i \vec \sigma \cdot \vec D \chi - \lambda \vec \tau \cdot \vec \Phi i \sigma^2 i \tau^2 \chi^\star
+ m i \sigma^2 i \tau^2 \chi^\star~~.
\ee
Here we have assumed $m^\dagger= + m$, and more specifically
\be
\label{basischoice}
\lambda = \( \begin{matrix} \lambda_1 & 0 \cr 0 & \lambda_2
\end{matrix}\)
~~~~~
m = i \( \begin{matrix}
 0 & \mu  \cr - \mu  & 0 \end{matrix}\)
\ee
with $\lambda_{1,2}, \mu$ real and positive.
The reality of $\mu$
(which implies that the Dirac mass matrix is hermitian)
is not fully general; we return to this point anon.

Following \cite{Jackiw:1975fn}, let
\be \chi_{\alpha a I} = \CM_{\alpha b I} i \tau^2_{ba} = \( \delta_{\alpha b} g_I + \sigma^i_{\alpha b} g^i_I \) \epsilon_{ba}
~.\ee
This decomposition incorporates the breaking of $\text{SU}(2)_{\rm gauge} \times \text{SU}(2)_{\rm spin}$
and decomposes $\chi \in (2,2)$ into irreps of the unbroken SU$(2)$.
It reduces the Dirac equation to the two equations:
\bea 0 &=& i \vec  \nabla g
- 2 i \hat r Ag
- 3 \vec g \times \hat r A
+  \lambda^\dagger g^\star \phi \hat r
\cr
&&
-   m^\dagger \vec g^\star
+ i \vec \nabla \times \vec g
+ \lambda \phi \hat r \times \vec g^\star
\cr
\cr
\label{eq:first} 0 &=& i \vec \nabla \cdot \vec g
+ 2 i \vec g \cdot \hat r A
-  \lambda^\dagger \vec g^\star \cdot \hat r \phi +  m^\dagger g^\star
~.\eea
The last term in \eqref{eq:first} forces us to include a nonzero $\vec g$ when $m \neq 0$.
The equations for $g, \vec g$ are (not too surprisingly) similar to
\cite{Jackiw:1975fn} equation A7a, b with extra terms coming from the Dirac mass.

We make an ansatz of the form $\vec g = \hat r g_r(r) $.
This eliminates the curl terms in the Dirac equation, leaving
\bea 0 &=& i \vec  \nabla g  - 2 i \hat r Ag -  \lambda^\dagger g^\star \phi \hat r -   m^\dagger \vec g^\star
\cr\cr
0 &=& i \vec \nabla \cdot \vec g  + 2 i  A \vec g \cdot \hat r +  \lambda^\dagger \vec g^\star \cdot \hat r \phi + m^\dagger g^\star ~.\eea

We choose the phases of $g, g_r$ so that $ i g = g^\star$
\be \label{eq:rephase} g \equiv \alpha h, ~~ g_r \equiv \alpha^{-1} h_r , ~~~~ \alpha \equiv e^{- \pi i /4}. \ee
The Dirac equation becomes
\begin{align}
 0 &=  \vec  \nabla h  - 2 \hat r A h +  \lambda^\dagger h \phi \hat r +  \mu \epsilon \vec h
\label{zmeqn1} \\
0 &=  \vec \nabla \cdot \vec h  - 2 A \vec h +  \lambda^\dagger \vec h \cdot \hat r \phi - \mu \epsilon h
~.
\label{zmeqn2}
\end{align}
In \eqref{zmeqn1},\eqref{zmeqn2}, all complex phases are explicit.
With the assumption \eqref{basischoice}, we have $\lambda^\dagger = \lambda$ is diagonal.
The  $\epsilon$ symbol acts on the $IJ$ flavor indices, and is the only thing which does.

The particular solution of \eqref{zmeqn2} for $\vec h$ given the source $h$ is:
\be
\label{eq:vecsol}
h_r = +\mu \epsilon r^{-2} e^{-\tilde H} \int^r s^2 e^{\tilde H} h(s)
\ee
where $ \tilde H \equiv  \int^r \( \tilde \lambda \phi - 2 A \) $
and $\tilde \lambda$ has the property that
\be \tilde \lambda m = m \lambda, \ee
which in turn requires
\be
\tilde \lambda = \( \begin{matrix} \lambda_2 & 0 \cr 0 & \lambda_1
\end{matrix}\) ~~~.
\ee
Plugging the solution \eqref{eq:vecsol} into \eqref{zmeqn1}
(and remembering that $\epsilon^2 = -1$) gives
\be \partial_r h + ( \lambda \phi - 2 A ) h = + \mu^2 r^{-2} e^{\tilde H} \int^r ds s^2 e^{-\tilde H} h ~.\ee
Substituting $ h = e^{-H} \gamma$ with 
$ H \equiv  \int^r \( \lambda \phi - 2 A \) $
gives
\be
\label{eq:IDE}
r^2 \partial_r \gamma = \mu^2 e^{H-\tilde H} \int^r ds s^2 e^{ \tilde H- H} \gamma~.
\ee

Differentiating \eqref{eq:IDE} (and thereby introducing an extra integration constant) gives
the linear second-order ODE for $\gamma$:
\be\label{eq:helm}
 r^{-2} \partial_r e^{\tilde H-H} r^2 \partial_r \gamma = \mu^2 e^{ \tilde H-H} \gamma~~.
\ee
We know the asymptotic behavior of the solutions at large and small $r$.
At small $r$, $H, \tilde H \to 0$, and the equation \eqref{eq:helm} reduces to the Helmholtz equation
\be \nabla^2 \gamma \equiv r^{-2} \partial_r r^2 \partial_r \gamma = \mu^2  \gamma
\ee
whose solutions are
\be
\label{eq:helmholzsol}
 \gamma \buildrel{r\to0}\over{\approx}c^{(-)} \frac{e^{-\mu r}}{r} + c^{(+)} \frac{e^{+ \mu r}}{r}.\ee
The combination of these solutions which also solves the integrodifferential
equation
\eqref{eq:IDE}
 in the small-$r$ regime has $c^{(+)} = - c^{(-)} \equiv - c$:
 \be
 \gamma \buildrel{r\to0}\over{\approx} {c\over r}\( e^{-\mu r} - e^{+ \mu r} \) ~.\ee
Note that only the combination
$  \tilde H - H = \lambda_T  \Ione \int^r \phi $
(where $ \Ione_{IJ} \equiv \delta_{IJ}, \lambda_T =  \lambda_1-\lambda_2$)
enters this equation.
We emphasize that there is one such solution for each value of the flavor index $I=1,2$,
labelled by a real integration constant $c_I$:
\be
\label{eq:sols}
 \gamma_I \buildrel{r\to0}\over{\approx} {c_I\over r}\( e^{-\mu r} - e^{+ \mu r} \) ~.\ee
In the special case where the eigenvalues of $\lambda$ are degenerate, $\lambda_T=0$, the equation for $\gamma$
is exactly the Helmholtz equation.  In this case, the solution
\eqref{eq:sols} is exact.

At large $r$, $ \phi(r)\buildrel{r\to\infty}\over{\approx} v $, and $A(r) \buildrel{r\to\infty}\over{\approx} \frac{a_0 }{r} $.
Therefore
\be H  \buildrel{r\to\infty}\over{\approx}
\( \begin{matrix} \lambda_1 v r  & 0 \cr 0 & \lambda_2 v r
\end{matrix}
\)
~, ~~~~ \tilde H -H \buildrel{r\to0}\over{\approx}  \lambda_T v r \Ione.
~~~\ee
To discuss the normalizibility of the solutions
at large $r$, we distinguish various parameter regimes.
\begin{itemize}
\item{}
If $\mu=0$, both solutions $\gamma_I$
in \eqref{eq:sols}
are normalizible for all $\lambda_{1,2}$.
Varying the signs or phases of $\lambda_{1,2}$ is innocuous;
it merely changes the overall phase of the zeromode solution
and can be absorbed in a field redefinition.

\item{} For small $\mu$,
\be \mu < \half  | \lambda_1 - \lambda_2 | v ~~,\ee
both zeromodes are still normalizible.

\item
Since the zeromode wavefunctions involve products of exponentials
of the form
$e^{\mu r} e^{- \lambda v r}$,
one might have thought (pantingly) that one zeromode would become
non-normalizable, \eg for $\mu$ in between the two Yukawa-induced fermion masses
\be\label{eq:ring}    \lambda_1v  < \mu <  \lambda_2  v .\ee
This hope is not realized -- there is no change in the normalizability of the 
modes at $\mu = \lambda_1v$.  

\item For  $\mu$ larger than the geometric mean of the fermion masses,
\be
\label{eq:normble}  \sqrt {\lambda_1 \lambda_2} v  < \mu ~~, \ee
{\it both} modes are non-normalizable.  There is no value of the parameters 
for which an odd number of Majorana modes are normalizable.

\end{itemize}

\begin{figure}[h] \begin{center}
\includegraphics[height=120pt]{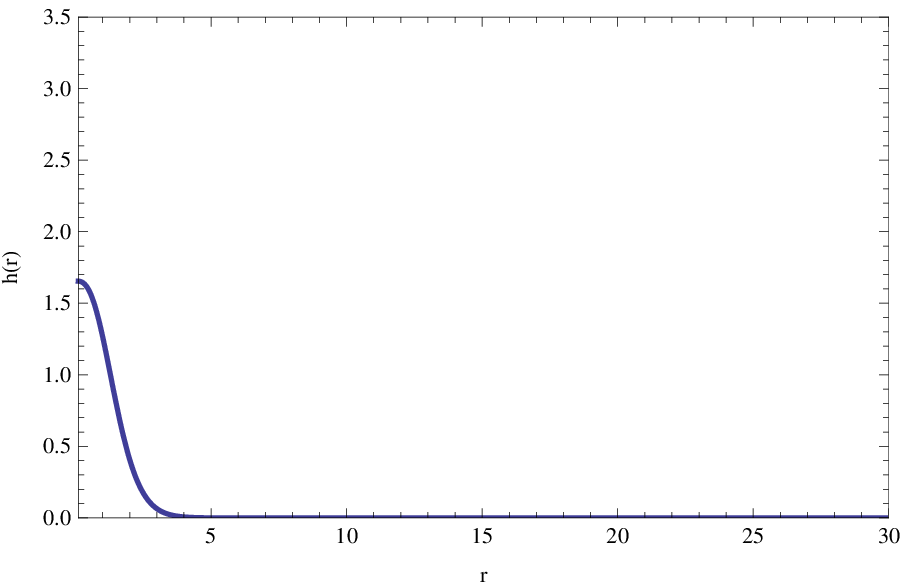}
\includegraphics[height=120pt]{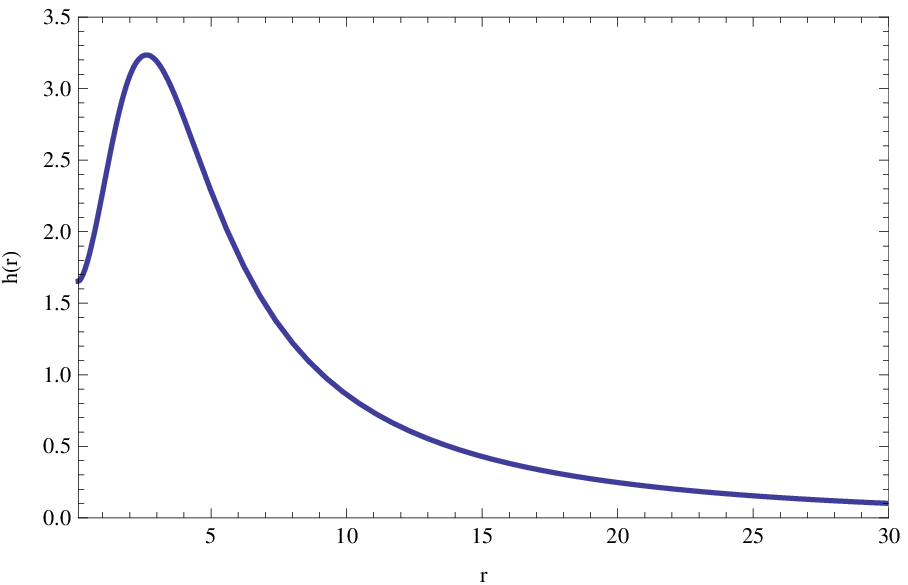}
\end{center}
\caption{Top: The profile of the zeromode solution for $\mu=0$.
Bottom: One of the profiles in the parameter range \eqref{eq:ring},
exhibiting the ring-like structure.
\label{fig:ph1}
}
\end{figure}

It is interesting to note that we are free to tune the effective sizes of the two real zero modes independently of each other.  By adjusting $\lambda_1$ and $\lambda_2$ we can produce a shell-like configuration of zero modes.  
More precisely, by making one of the fermion masses very heavy, we can arrange
(in the parameter regime \eqref{eq:ring})
for only one zero mode to have a sizable wavefunction until very close to the monopole core, as 
shown in FIG.~\ref{fig:ph1}.  Whether this separation of scales could in principle allow for interesting physical effects is not clear to us.

Note that the variation of the bulk fermion spectrum with $\mu$ corroborates
the understanding of the normalizability properties of the zeromodes presented above.
The product of the bulk fermion masses is the determinant of
the fermion mass matrix
\be
M = v \lambda \otimes \tau^3 + m \otimes \Ione
=  \( \begin{matrix} \lambda_1v  & i \mu & 0 & 0 \cr
- i \mu & \lambda_2 v & 0 & 0 \cr
0 & 0 & - \lambda_1 v & i \mu  \cr
0 & 0 & - i \mu & - \lambda_2 v   \cr \end{matrix} \)
\ee
which is
\be \label{eq:marginal}
\det M = \( \lambda_1 \lambda_2 v^2 - \mu^2 \) ^2~~.
\ee
Comparing 
\eqref{eq:marginal} to the condition for
normalizability of the zeromodes on the monopole, 
\eqref{eq:normble}, 
we see that 
precisely when the zeromodes become marginally normalizable, there is a massless fermion in the bulk.
As $\mu$ is increased the zeromodes leak out of the monopole core and join the bulk states.

If $m$ is not Hermitian, any rephasing analogous to
\eqref{eq:rephase} produces overconstraining equations:
the solutions are forced to have nonzero energy. As we discussed above, a CP symmetry can enforce hermiticity of $m$.

\subsection{Non-relativistic mass}
\label{sec:NRmass}

The non-relativistic mass $m \chi^\dagger \chi$ appears in the Dirac equation
in precisely the same way as the energy.
In fact, this term is nothing but a chemical potential for the chiral symmetry, and thus it clearly breaks Lorentz invariance while preserving rotational symmetry.  As the full chiral symmetry is anomalous, this term produces a finite density of fermions carrying a non-conserved charge.  This symmetry is also explicitly broken by the scalar coupling, and so even without the anomaly the chiral symmetry is broken as in a superconductor.  As the fermion spectrum remains fully gapped in the presence of the scalar coupling, we expect that the non-relativistic mass does not seriously affect the zero-mode spectrum.  This must be true in the ungauged theory of a single Weyl doublet coupled to a scalar field in the adjoint, as such a theory has only a single Majorana mode on a hedgehog that cannot pair and disappear.

\section{5d realization}
\label{sec:5d}

Consider SU$(2)$ Yang-Mills theory
in 4+1 dimensions
with a 5d Dirac fermion
in the doublet representation, and
an adjoint scalar in its condensed phase.
Identify the fourth spatial dimension $y \simeq y + 2 \pi R$.
Consider a kink-antikink configuration of the 5d Dirac mass $M(y)$ of the fermion,
with the kink and antikink on opposite sides of the circle,
that is
\be M = +m, y \in (0, \pi R), ~~~ M= - m, y \in (\pi R, 2\pi R).\ee
The kink and antikink each support a 4d massless Weyl fermion (for a useful review, see \cite{Kaplan:2009yg}).  We can arrange for the 4d coupling to the scalar field that we have been considering
by using the fact that spinor representations in 5d are also pseudoreal (the Lorentz group is equivalent to a symplectic group which has a real invariant form).  Using the 4d chiral basis, a 5d spinor may be written $\Psi = [ \psi_L ,\, \psi_R ]^T$.  The combination $\psi_L^T i \sigma^2 \psi_L + \psi_R^T i \sigma^2 \psi_R$ is manifestly invariant under 4d Lorentz transformations.  The extra four transformations in the 5d Lorentz group, generated by $[\gamma^4, \gamma^\mu]$, act infinitesimally like
$\delta \psi_L = \epsilon \sigma^\mu \psi_R$ and $\delta \psi_R = -  \epsilon \overline{\sigma}^\mu \psi_L$. The invariance under 5d Lorentz transformations then follows from the identity $(\sigma^\mu)^T i \sigma^2 + i \sigma^2 \overline{\sigma}^\mu  = 0$.
(That is, the symplectic invariant of $SO(4,1)$ is $\Omega \equiv \gamma^1 \gamma^3$.)
The full coupling is then
$$ \psi_L^T i \sigma^2 i \tau^2 \Phi \psi_L + \psi_R^T i \sigma^2 \tau^2 \Phi \psi_R
= \Psi^T i \tau^2 \Phi \Omega \Psi. $$

We would like to view this model in analogy with
lattice realizations of a single 2+1-dimensional Dirac fermion
on the boundary of a 3+1 dimensional lattice.
The extra dimension allows one to evade
the lattice doubling no-go theorems \cite{Nielsen:1980rz,Nielsen:1981xu,Nielsen:1981hk}.
The Witten anomaly seems to be cancelled by inflow from the bulk.
The precise meaning of the previous sentence
could be clarified given an explicit expression for the WZW functional $\Gamma[A, \Phi]$.

This model is unsatisfactory in at least three ways.
First, its five-dimensional nature may make it hard to realize in the laboratory.
Secondly, 5d Yang-Mills theory is not asymptotically free and must be
completed at short distances somehow
(string theory gives interesting ways to do this, \eg \cite{Seiberg:1996bd};
this model can also be latticized).
Thirdly, if we allow the profile of the mass to fluctuate,
the kink and antikink can annihilate each other.
Nevertheless, the model is instructive.

The model has many mass scales:
the W-boson mass, $M_W$,
the Kaluza-Klein scale $R^{-1}$,
the Dirac mass $m$,
the inverse thickness of the kink,
and an extreme UV cutoff above
which the gauge theory succumbs to higher-energy physics.
The last two we suppose to be inaccessibly high.

At energies $E \ll 1/R$, this model reduces
to the two-doublet theory studied in the previous section.

Having added an extra spatial dimension, monopoles
(whose topological charge is characterized by a non-trivial element of $\pi_2(S^2)$)
now become string-like objects which we refer to as `monopole strings'.
Consider a monopole string winding around the fifth dimension at
some point in 3d space,
$ \vec r  = 0 $.
From a 4d point of view, this appears to be a magnetic monopole.  This follows from the fact that the monopole string current $J^{\mu \nu}_M$ sources the 5d U$(1)$ field strength via $dF = J_M$.
Where the vanishing loci of the order parameter $\Phi$
and the 5d Dirac mass intersect,
the 5d Dirac equation will support localized Majorana zeromodes.

This model demonstrates that the two Majorana modes
need not pair up.  Here their wavefunctions are separated
in the extra dimension.
In the regime $ m \gg R^{-1} $,
their overlap is exponentially small.

To illustrate the physics of this 5d construction, we consider a configuration of four monopole strings each parallel to the compact direction and wrapping once around it.  Each of the $i=1,...,4$ monopole strings intersects each of the $a = 1,2$ domain walls once for a total of $4*2 = 8$ Majorana modes that we label $\gamma_{i a}$.  These operators satisfy the algebra $\{ \gamma_{i a}, \gamma_{j b} \} = 2 \delta_{i j} \delta_{a b}$ and are real $\gamma^+_{i a} = \gamma_{i a}$.  From the point of view of 4d physics, a natural basis for this space of states comes by forming complex fermions $c_i = \frac{\gamma_{i 1} + i \gamma_{i 2}}{2}$ made from Majoranas at the same point in the non-compact directions.  Using these fermion operators we can build a space of $2^4$ states which further subdivides into an $8$ dimensional subspace of even fermion parity and an $8$ dimensional subspace of odd fermion parity.

Three important questions must now be answered.  First, what states can be produced by creation of such a system from the vacuum state (or any other state without such a configuration of monopole strings)?  Second, what operations on the monopole strings can be carried out without large energy cost?  Third, what decoherence free superpositions are possible?

The first question has two immediate answers.  The simplest local (from the 4d point of view) vacuum-like state is the state annihilated by all the $c_i$ defined above.  The Majorana modes $\gamma_{i 1}$ and $\gamma_{i 2}$ can be viewed as the ends of a ``quantum wire" as in \cite{Kitaev:2000} and it is quite natural from the 4d point of view to
pair up these Majoranas.  The second immediate answer comes from thinking about the creation process by which such a monopole string configuration could be formed.  For example, we could take monopoles $1$ and $4$ to have magnetic charge $1$ and monopoles $2$ and $3$ to have magnetic charge $-1$.  Then we could pair create $1$,$2$ and $3$,$4$ from the vacuum state.  With this process in mind, and
remembering that the Majorana wavefunction overlap in the compact direction can be made exponentially small, a natural initial state would be that state annihilated by complex fermions formed from Majoranas on neighboring monopoles (independently for each domain wall).  This state also has even fermion parity but is not equal to the state annihilated by all the $c_i$.

As for low energy operations, we must at least require no macroscopic stretching of the monopole strings beyond that required to have the monopole wrap the compact direction.
 If $\CT$ is the monopole tension, then the mass of the monopole string is $2 \pi \CT R$.
  In order to perform operations on the zero mode Hilbert space, we would like to entertain motions of the monopole strings.
    However, we must move an entire monopole string at once in order to avoid a large energy cost associated with stretching the monopole string.
This always means exchanging pairs (coming from the two domain walls) of Majorana modes.
For example,
consider exchanging monopole
strings $2$ and $3$ in the
configuration above.
  This implements the operator
$$U_{2 \rightleftharpoons 3} = \exp{\(\frac{\pi}{4} \(\gamma_{2 1} \gamma_{3 1} + \gamma_{2 2} \gamma_{3 2}\) \)} ,$$
but this operation can be reexpressed in terms of the $c_i$ as
$$U_{2 \rightleftharpoons 3}=\exp{\(\frac{\pi}{2}\(c_2 c_3^+ + c_2^+ c_3\)\)}~~.$$

This operator acts trivially on states with $c_2^+ c_2 = c_3^+ c_3$ and exchanges pairs of states with $c_2^+ c_2 = 1 - c_3^+c_3 \,\text{mod}\, 2$.  In other words, it simply moves around local fermions from the 4d perspective.  Note also that we have not included the dynamics of the gauge field during this exchange process.  We note in passing that there is interesting physics associated with the dynamics of the gauge field, particularly the role of instantons, for example, an instanton localized along a line in 5d spacetime describes a conduit via which fermions tunnel from one wall to the other.  Since the 4d local basis effectively stacks the two Weyl fermions on top of each other, the physics should be qualitatively similar to the case of a single Dirac fermion in 4d discussed above.  In particular, once the gauge field motions are included, we find that the states built from the $c_i$ operators are actually all bosonic because of the extra angular momentum coming from the gauge field.

Finally, what about decoherence free superpositions?  The 4d local basis $c_i$ seems naively decoherence free, but another regime is possible where the smallest scale is $R^{-1}$.  In this regime, the Abelian gauge field resulting from the Higgsing of SU$(2)$ looks five dimensional and may even decohere the fermions in the 4d local basis generated by the $c_i$.  However, in this case we are faced with the question: decohere to what?  There seems to be no local basis once the gauge field is allowed to fluctuate in the 5th dimension.  There is also no superconductivity to justify forming decoherence free superpositions of different charge states.  In fact, there is an even simpler configuration that can cause concern.  Consider a single monopole string forming a closed loop
which does not wrap the extra dimension but still punctures one of the domain walls twice.  Now this configuration may cost a lot of energy and be unstable, but assuming we could hold the monopole string in place, we appear to have two Majorana modes on a single domain wall but again with no obvious local basis to decohere into.  We are again faced with the question: decohere to what?

To resolve these issues, we need to bring in a thus-far neglected piece of the puzzle.  In 4d the SU$(2)$ monopole has a collective coordinate, a rotor degree of freedom corresponding to the unbroken U$(1)$ charge.  The excitations of the rotor generate the familiar dyon states of the monopole.  In the 5d model we have a new complication: instead of a single quantum mechanical rotor, we are faced with a rotor degree of freedom for each point on the monopole string.  Thus the monopole string supports a finite-size realization of the $1+1$ dimensional XY model, a $c=1$ conformal field theory.  These gapless degrees of freedom can significantly affect the physics.  Charge will be dynamically screened by the gapless rotor degrees of freedom living in the monopole string core.  In the parameter regime where the compact radius is large, we have 5d U$(1)$ gauge theory, and the only configurations of the Majorana zero modes that remain decoherence-free are those connected by a single rotor string, and they are still linearly
confined by the monopole string tension.  Thus for strings wrapping the compact direction the decoherence free subspace is always the 4d local basis and we recover the low energy physics of the Dirac fermion coupled to a scalar in 4d as we must. We can also consider monopole strings as above that intercept only one domain wall, but here the majorana zero modes are bound by the monopole string stretching between them, the same string that screens their gauge charge. 

\section{Conclusions and general arguments}
Despite a promising attempt, we have
not found a consistent field theory with
Majorana monopoles that are not linearly confined.
We would like to argue that this conclusion is general, and will do so from a variety of points of view.

\subsection{Monopole configuration space}
If we had found a consistent gauge theory with unpaired Majorana operators on the cores of monopoles, we would have been in serious trouble.  Indeed, the fundamental group of the bare $N$-monopole configuration space is precisely $S_N$, and we know that this group has no interesting non-local representations \cite{Leinaas:1977fm,Doplicher:1971wk,Doplicher:1973at,deligne:tcat}\footnote{Deligne's theorem \cite{deligne:tcat} proves that replacing the braid group by the symmetric group gives ``local" theories called $\mbox{Rep}(G,\mu)$.  We have bosons ($\mu=1$) or fermions ($\mu=-1$) with a local ``internal" symmetry $G$.}.  The existence of extended magnetic field lines does not help since the static magnetic field configuration is completely specified by the positions of the monopoles via the magnetic Gauss law.  One might have hoped that the Dirac string, which is the remnant of the
ribbon
that proved so essential in the ungauged theory \cite{Freedman:2010ak}, could play a similar role here.  However, this string is unphysical as its position can be moved using gauge transformations.  For example, in lattice U$(1)$ gauge theory the Dirac string is completely meaningless and undefined.  Thus the only remaining possibility is the existence of some subtle topological information encoded in the existence of the Dirac string (but not its precise position) in certain UV completions of U$(1)$ gauge theory.  We can find no such data and although we do not prove it cannot be found, we regard this possibility as quite remote.  The main point is simply that the configuration of monopoles in a Coulomb phase is insufficient to support non-Abelian particles.  One would have to add extra data beyond the monopole positions in any model that realized non-Abelian particle-like excitations.

\subsection{Callias index and anomaly}

Here we make a precise connection between the Majorana number mod two and the Witten anomaly.  Roughly, we can relate the Witten anomaly to the chiral anomaly mod two;
in turn we can relate the chiral anomaly mod two to
the Majorana number of the monopole.

In a theory with a Witten anomaly,
a chiral rotation by $\pi$ is an element of the gauge group \cite{Goldstone:1984, deAlwis:1985uj},
{\it i.e.} $(-1)^F$
acts in precisely the same way as
a gauge rotation $e^{i \pi \tau_3}$ for some gauge generator $\tau_3$.
One way to think about this statement is
that there are no gauge-neutral excitations
which carry unit fermion number;
this means that the fermion number
and the gauge charge are the same mod two.
The chiral anomaly mod two
is therefore in fact a gauge anomaly
\cite{Goldstone:1984, deAlwis:1985uj}.
In the Witten-anomalous theory,
the chiral anomaly --  {\it i.e.} the fact that an instanton violates the chiral charge
by one unit (destroys a RH fermion {\it or} creates a LH fermion) --
means that the instanton must also violate the gauge symmetry
(despite the fact that there is no local gauge anomaly)\footnote
{We note in passing that the fermion-number violation by instantons
seems to be a {\it symptom} of the Witten anomaly,
rather than an equivalent statement.
We say this for the following reason.
Recently \cite{Pantev:2005rh, Hellerman:2006zs, Seiberg:2010qd} it has been argued 
that
it is possible to modify gauge theories by restricting the instanton sum,
for example to even instanton numbers.
This would make the partition function $\pi$-periodic
in $\theta$ solve the obstruction given in eqns (18, 19) of \cite{deAlwis:1985uj}.
However, applying the path-integral method for accomplishing
this modification given in \cite{Seiberg:2010qd}
to a Witten-anomalous theory does not
change the fact that the fermion determininant
faithfully represents $\pi_4(\text{SU}(2))$,
and therefore does not prevent the gauge field path integral from vanishing.}.

The chiral anomaly mod two in turn is related
by (a generalization of) Callias' index theorem
\cite{Callias:1977kg}
to
the number of {\it real} fermion zeromodes of
the monopole.
The result proved by Callias counts the index of a complex-linear Dirac operator;
this is the number of {\it complex} zeromodes weighted by some version of chirality.
Because of the coupling to the Higgs field $\Phi$, our Dirac operator is only
real-linear, and we wish to count its {\it real} zeromodes
(in a monopole background), mod two.
This kind of zero mode counting has been considered in
 \cite{nishida:2010},
and they concluded that the Chern number indeed counts the Majorana number mod two.  Thus it seems that within the setup of microscopic fermions coupled to an SU$(2)$ gauge field and an adjoint scalar, the existence of an unpaired Majorana zero mode in the ungauged theory is unavoidably related to the presence of the Witten anomaly in the gauge theory.

\subsection{More arguments from low energy}
More generally, we could ask if deconfinement is possible via the disordering route mentioned in the introduction.  This scenario has at least two problems.  First, as we argued above, the configuration space of monopoles is too trivial to support non-Abelian particles.
It appears we must gauge away the
ribbon data
or disorder it away.  Second, given the unbroken global SU$(2)$ symmetry in the disordered phase, the quantum numbers of local excitations should be consistent with the unbroken symmetry.  It is hard to see how we can build a sensible real zero mode without doing violence to the SU$(2)$ group structure.

This question can be addressed in more detail using the slave particle techniques which have been
developed for the study of spin liquids
(\ie\ disordered groundstates of quantum spin systems).
In the disording scheme described in the introduction, we write the order parameter in terms of bosons $z_\alpha$ as $n^i = z^+ \tau^i z$.  In a fractionalized phase with unbroken SU$(2)$ symmetry the doublet fermions will be screened by $z$ and become SU$(2)$-neutral; however,
the resulting SU$(2)$-singlet fermions
will carry an internal U$(1)$ gauge charge.  As before, for any hope of success we must disorder the $n$ field without condensing hedgehogs.  Assuming we can do this, hedgehogs will become monopoles of the emergent U$(1)$ gauge field.  It appears difficult to form the necessary decoherence free superpositions of fermions charged under the internal U$(1)$ to produce Majorana zero modes on the monopole cores.  We also still have the problem of the monopole configuration space.  Thus we argue that such a phase is either impossible or the number of Majorana zero modes on the monopole changes across the phase transition.

It is possible for the $U(1)$ gauge symmetry
to be found in a Higgs phase;
in this case the
Majorana solitons are monopoles in a superconductor which again are linearly confined
by magnetic flux tubes, and it is perfectly consistent to have
localized states of indefinite charge.

We can consider other possibilities, where there is no $U(1)$ gauge symmetry
at any energy scale.
For example, one could try to decompose the order parameter as $n^i = b^T i \tau^2 \tau^i b$ with $b$ a two component complex doublet of bosons.  Now the disorded phase will only have an emergent $Z_2$ gauge field, but the original order parameter has an extra U$(1)$ symmetry associated with $b \rightarrow e^{i \theta} b$ (whereas the SU$(2)$ transformation is $b \rightarrow e^{i \theta \tau^3} b$.  In other words, $n$ must be complex.  Even if we break this symmetry in the Hamiltonian we can still unwind hedgehog configurations using the extra scalar degrees of freedom.  This is to be expected since the hedgehog would have turned into a localized object in the $Z_2$ gauge theory, but there is no local object in such a theory in $3+1$ dimensions (the vortex from $2+1$ is now a vortex line in $3+1$).

The possibility remains that a 3+1-dimensional {\it lattice} model
exists with deconfined Majorana monopoles,
{\it i.e.} that the continuum limit (our starting point) fails to capture some
crucial element.
Certain kinds of lattice models that begin with Majorana fermions may, not surprisingly, more easily produce Majorana excitations.
If these models cannot be realized with a ``proper" regularization involving only complex fermions coupled to superconductivity, then we are tempted to regard them as too artificial.  We can easily design a network of Kitaev quantum wires in three spatial dimensions that reproduce the topological aspects of the Teo-Kane model, however there is no SU$(2)$ symmetry (it is reduced to a discrete subgroup) and the confinement is still linear.  Without the full SU$(2)$ symmetry we cannot gauge the model.  Furthermore, there can be no 4d lattice realization of the Teo-Kane model with full SU$(2)$ symmetry since such a lattice model, when attached to the surface of the 5d model above, would produce a trivial surface.  Put differently, if such a lattice model did exist it could be trivially gauged and we would face the Witten anomalous gauge theory again.

We started from a desire to produce deconfined non-Abelian particle-like excitations in $3+1$ dimensions.  Specifically, we were interested in localized objects displaying what could be called Majorana statistics.  The perhaps simplest route to deconfinement led to an anomalous gauge theory.  In attempting to cure the anomaly, we found repeatedly that deconfinement requires the number of Majorana zero modes to be even, giving ordinary statistics. We have made many attempts: high energy fermionic matter, extra dimensions, disordered phases exhibiting emergent gauge fields, but none led to deconfined non-Abelian particles.  This is all completely consistent with general expectations about the nature of particle excitations in three dimensional space.  We conclude with a few comments for future work.  We always find linear confinement, but this may not be the most general situation. For example, we can argue that gauging only a subgroup of the full SU$(2)$ symmetry still leaves linear confinement intact. So how strongly bound must such non-Abelian particles be in general?  Finally, there remains the prospect that with the right low energy data, deconfined non-Abelian particles would be possible.  Although we have ruled out many promising paths to this goal, it would be very exciting to see such a possibility realized elsewhere.

{\bf Acknowledgements}

JM thanks Frank Wilczek for sharing
ideas on a model similar to the gauged Teo-Kane model
in January 2009.
We would like to thank Andrea Allais, Jae Hoon Lee, Chetan Nayak, Zhenghan Wang, Mike Mulligan, T.~Senthil
for useful discussions and comments.  We thank T.~Senthil for collaboration in the initial stages of this work.
3D graphic from the Logo Company \cite{logo}. The work of JM is supported in part by
funds provided by the U.S. Department of Energy
(D.O.E.) under cooperative research agreement DE-FG0205ER41360,
and in part by the Alfred P. Sloan Foundation.

\bibliography{majorana_monopole}
\end{document}